\documentclass[a4paper,prb,twocolumn,superscriptaddress,nofootinbib]{revtex4-2}
\pdfoutput=1

\usepackage{amsmath,amsfonts,color,bm,graphicx,mathtools,datetime}

\newcommand{\ket}[1]{\left\vert#1\right\rangle}
\newcommand{\bra}[1]{\left\langle#1\right\vert}

\begin{document}

\title{Topological Spin Liquids: Robustness under perturbations}

\author{Mohsin Iqbal}
\affiliation{Max Planck Institute of Quantum Optics,
Hans-Kopfermann-Str.~1, 85748 Garching, Germany}
\affiliation{Munich Center for Quantum Science and Technology, 
Schellingstr.\ 4, 80799 M\"unchen, Germany} 
\author{Helena Casademunt}
\affiliation{Department of Physics, Princeton University, Princeton NJ,
United States}
\affiliation{Department of Physics, Harvard University, Cambridge MA,
United States}
\author{Norbert Schuch}
\affiliation{Max Planck Institute of Quantum Optics,
Hans-Kopfermann-Str.~1, 85748 Garching, Germany}
\affiliation{Munich Center for Quantum Science and Technology, 
Schellingstr.\ 4, 80799 M\"unchen, Germany}

\begin{abstract}
We study the robustness of the paradigmatic kagome Resonating Valence Bond
(RVB) spin liquid and its orthogonal version, the quantum dimer model. The
non-orthogonality of singlets in the RVB model and the induced finite
length scale not only makes it difficult to analyze, but can also
significantly affect its physics, such as how much noise resilience
it exhibits.  Surprisingly, we find that this is not the case: The amount
of perturbations which the RVB spin liquid can tolerate is not affected by
the finite correlation length, making the dimer model a viable model for
studying RVB physics under perturbations.  Remarkably, we find that this
is a universal phenomenon protected by symmetries: First, the dominant
correlations in the RVB are spinon correlations, making the state robust
against doping with visons. Second, reflection symmetry stabilizes the
spin liquid against doping with spinons, by forbidding mixing of the
initially dominant correlations with those which lead to the breakdown of
topological order.  \end{abstract}

\maketitle

\section{Introduction}

Topological spin liquids (TSL) are exotic phases of matter where a system
does not order magnetically despite strong antiferromagnetic interactions,
but rather topologically, i.e.\ in its global entanglement.  The interest
in these systems stems from their unconventional properties, such as
anyonic excitations with fractional charge and non-trivial statistics, and
a ground space protected by its global
entanglement~\cite{balents:spin-liquid-review-2010,savary:spin-liquids,knolle:spin-liquids}.
However, TSLs are notoriously difficult to identify, both in
theory and in experiment, as candidate systems often exhibit close
competition between a number of different phases. In order to robustly
realize these phases, it is therefore essential to understand how 
sensitively they react to perturbations which
can induce a breakdown of topological order.

The paradigmatic example of a spin liquid is the Resonating Valence Bond
(RVB) wavefunction on the kagome lattice, which consists of a
``resonating'' superposition of all possible ways to cover the lattice
with nearest-neighbor singlets~\cite{anderson:rvb,misguich:AFM-review}. 
It forms a physically motivated low-energy ansatz for Heisenberg-type
models, and appears as the exact ground state of a local 
model with topological
order~\cite{schuch:rvb-kagome,zhou:rvb-parent-onestar}.
However, the non-orthogonality of different singlet configurations makes
RVB models hard to analyze. To mitigate this difficulty, dimer models have
been studied instead, where different singlet configurations are
taken to be orthogonal~\cite{rokhsar:dimer-models}.  The resulting kagome dimer
model is an RG fixed point and thus significantly easier to
analyze~\cite{misguich:dimer-kagome}.
However, it is unclear to what extent results derived for the dimer model
still apply to the RVB state, where the non-orthogonality of singlets
induces a finite correlation length, making it
doubtful whether the
robustness of the RVB state can be understood from studies performed on
the dimer model.

In this paper, we study and compare how sensitively the RVB 
spin liquid and the quantum dimer model
react to noise, and which level of perturbations they can tolerate before their
topological order breaks down.  We consider both magnetic fields and
lattice anisotropies, corresponding to doping with the two elementary
topological excitations: spinons and visons. We find, rather surprisingly, 
that in both cases the RVB model exhibits essentially the same
stability as the dimer model despite its non-zero correlation length. 
This suggests that the dimer model is more accurate in modeling spin
liquid physics under perturbations than one might have naively assumed.

To understand the mechanism behind this unexpected result and its
range of applicability, we microscopically analyze the structure of anyon
correlations using tensor networks. Diverging anyon correlations
indicate a closing gap, driving a phase transition through anyon
condensation; the finite spinon correlations in the RVB would thus indeed
suggest a decreased robustness.  However, as our analysis reveals, there
is a \emph{universal} mechanism underlying the surprising
robustness of the RVB model:  It arises from symmetries which protect
specific correlations, and is thus independent of the specific
perturbation but rather a universal feature of the RVB spin
liquid.

Concretely, we find that the
non-orthogonality of singlets induces dominant spinon correlations without
a separate vison correlation scale; since vison doping only increases the
latter, the response is unaffected by the spinon length
scale.
The reason for the robustness to spinon doping is more subtle;
we assess it through a combination of analytical and numerical
study. 
%
%
It reveals that the spinon correlations exhibit a two-fold degeneracy in
addition to the spin doublet. It originates in the two different ways to
construct spinon correlations -- either through the overlap of two states with
one spinon each, separated by some distance $\ell$, or through the overlap of
the original doped RVB with a state with two additional spinons placed at
a separation $\ell$. We show that the reflection symmetry of the RVB rules
out any correlation between spinons of opposite spin, also at finite
doping.  This, together with
the symmetry of these overlaps under ket$\leftrightarrow$bra exchange, 
implies that the 4-fold multiplet splits under doping into sectors labelled
by (i) their spin $\pm\tfrac12$ and (ii) a fixed relative phase $\pm1$ between the
spinon at either position being in the ket or bra vector in  the overlap,
respectively.  As we show, this $\pm1$ phase label is protected by the
reflection symmetry of the lattice and thus stable to perturbations which
respect that symmetry.  
We find that 
the correlations which are initially enhanced by
magnetic fields and those which drive the phase transition live in
different sectors which are protected by the lattice symmetry; that 
explains why the presence of initial spinon correlations in the RVB state
has no effect on its robustness to magnetic fields.

\section{RVB and dimer model}

\begin{figure}[t]
\includegraphics[width=\columnwidth]{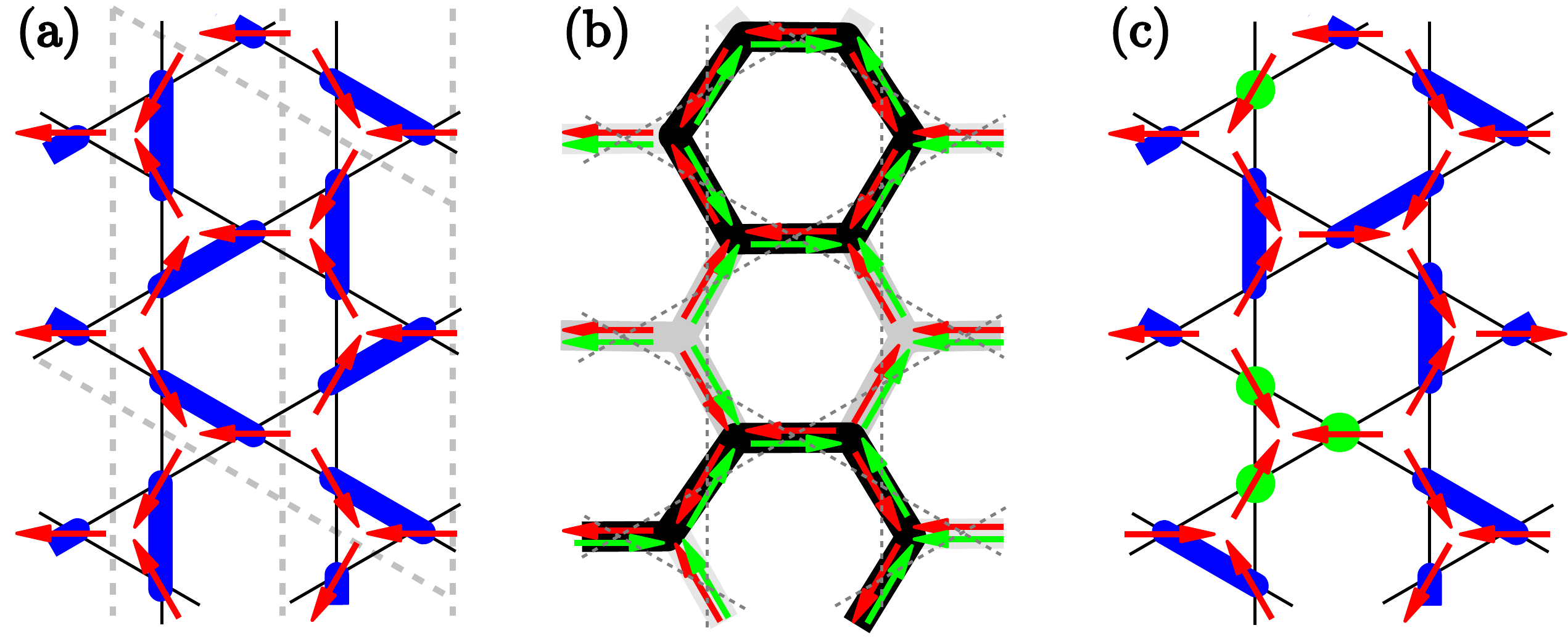}
\caption{\textbf{(a)} Dimer pattern (blue) and its
arrow representation (red); arrows point into the triangle with
the dimer. \textbf{(b)} The difference between any allowed arrow pattern
(green) and the reference pattern (red) is in one-to-one correspondence to
loop patterns. \textbf{(c)} Construction of ``dual tension'' doping,
cf.\ text.
}
\label{fig:rvb}
\end{figure}

The RVB state 
is constructed as follows.  First, we
define a dimer covering as a full covering of the
lattice with pairs of adjacent vertices, termed \emph{dimers}
(blue in Fig.~\ref{fig:rvb}a); 
we denote dimer coverings by $D$, and the set of all coverings
(with PBC) by $\mathcal D$.  Next, replace each dimer by a singlet
$\ket{\sigma}=\tfrac{1}{\sqrt{2}}(\ket{\uparrow\downarrow}-\ket{\downarrow\uparrow})$ 
(with counterclockwise orientation
around triangles), we call the resulting state $\ket{\bm{\sigma}(D)}$.  The
RVB wavefunction is then 
$\ket{\mathrm{RVB}}=\sum_{D\in\mathcal D} \ket{\bm{\sigma}(D)}$.
In studying the physics of RVB wavefunctions, so-called dimer
models $\ket{\mathrm{dimer}} = \sum_{D\in\mathcal D}\ket{D}$ are
frequently 
used, where the $\{\ket{D}\}_{D\in\mathcal D}$ define an
orthonormal basis~\cite{rokhsar:dimer-models}. Replacing singlet
configurations by orthogonal dimer configurations makes these models
easier to analyze, but can also affect their physics.
One way to explicitly
construct a dimer representation is to start from the RVB wavefunction and
attach arrows to each vertex~\cite{elser:rvb-arrow-representation}
which for any dimer configuration $D$ point into the
triangle where the adjacent dimer lies, Fig.~\ref{fig:rvb}a.
These arrows can be treated as quantum degrees of freedom or ``arrow qubits''
with basis states $\{\ket{a_\uparrow},\ket{a_\downarrow}\}$ (arrow pointing into either
of the two adjacent triangles); denoting the corresponding global arrow
configuration by $\ket{A(D)}$, we obtain a local representation
$\ket{\mathrm{dimer}} = \sum_{D\in\mathcal D}\ket{\bm\sigma(D)}\ket{A(D)}$
of the dimer model.
One advantage of this representation is that it allows to continuously
interpolate between the dimer model and the RVB state, by choosing a
non-orthogonal arrow basis
$\ket{a_{\uparrow/\downarrow}}=(1\pm\lambda)\ket{0}+(1\mp\lambda)\ket1$
 and tuning 
$\lambda\in[0;1]$. It can be proven that along the whole interpolation,
the system has a parent Hamiltonian with a $4$-fold degenerate ground
space with topological features, and numerical study shows
that the correlation length along the interpolation stays finite,
placing both models in the same (topological) phase without any
conventional order~\cite{schuch:rvb-kagome}.

The dimer model can be proven to be a topological fixed point model
using the arrow representation introduced above.  First, given
any classical configuration $\ket{A(D)}$, we can disentangle the singlets
$\ket{\bm{\sigma}(D)}$ by local unitaries (conditioned on the 
adjacent arrow qubits) and bring them to a fiducial state, leaving us
with a superposition $\sum_{A\in\mathcal A}\ket{A}$ of all allowed arrow
configurations $A\equiv A(D)$; these are precisely those with an even number of
inpointing arrows (a $\mathbb Z_2$
constraint)~\cite{elser:rvb-arrow-representation}. 
By fixing a ``reference configuration'' $A_0$ of arrows (and thus a
reference configuration $D_0$ of dimers, Fig.~\ref{fig:rvb}a), every arrow configuration
$A\in\mathcal A$ is characterized by those vertices where the arrows in $A$
differ from $A_0$.  These vertices satisfy a $\mathbb Z_2$ Gauss law on
each triangle and thus describe closed loops $L$ on the dual honeycomb
lattice (Fig.~\ref{fig:rvb}b); this establishes a one-to-one correspondence
between loop configurations $\mathcal L$ and arrow configurations
$\mathcal A$.  The dimer model is thus locally equivalent to an
equal-weight superposition of all loop configurations on the dual
honeycomb lattice, which is nothing but the topological $\mathbb Z_2$ Toric
Code model~\cite{kitaev:toriccode}. In fact, we can think of the dimer and
RVB model as being constructed from a
loop model to which we apply a sequence of local operations which replace
the loops by arrows, add singlets as prescribed by the arrows, and finally
(partially) erase the arrow pattern by applying 
$E_\lambda=\left(\begin{smallmatrix}1+\lambda & 1-\lambda\\
    1-\lambda & 1+\lambda \end{smallmatrix}\right)$
to each arrow qubit; note that while the first two steps are local
unitaries/isometries, the last step $E_\theta$ is a non-unitary
(``filtering'') operation which induces a finite correlation
length and in the limit $\lambda\to0$ becomes singular.

\section{Doping the RVB wavefunction}

Subjecting the RVB or dimer model to external fields induces doping with
elementary excitations: spinons or visons.  Spinons are obtained by
breaking up a singlet (or dimer) and replacing it by two separate spins
(w.l.o.g., two up-spins), which can subsequently separate due to a kinetic
term.
Visons, on
the other hand, correspond to a local disbalance in the relative weight of
different singlet (or dimer) configurations (equivalently, loop
configurations). 

In order to study how a finite density of excitations affects the
topological order in the RVB or dimer model, we extend the 
ansatz to include a tunable quasiparticle
doping. Let us start with vison doping: Here, we select a
reference dimer pattern $D_0$ (=arrow configuration $A_0$,
Fig.~\ref{fig:rvb}a), and
adiabatically increase the relative weight of
the reference configuration through a filtering 
$F_{\theta_v} = \openone-\theta_v\ket{\bar r}\bra{\bar r}$
(with $\ket{\bar r}$ the opposite of the reference state)
applied to each arrow qubit \emph{before} the application of $E_\lambda$.
This directly translates to a ``string tension''
$(\begin{smallmatrix}1&0\\0&1-\theta_v\end{smallmatrix})$
in the underlying loop model (defined relative to $D_0$) which suppresses
longer loops and gives rise to doping with magnetic (vison) excitations;
for the dimer point $\lambda=1$, the two doping models are unitarily
equivalent.

For spinon doping, we introduce two ansatzes.  The first -- termed ``dual
tension'' -- maps to electric excitations 
(broken loops). This corresponds to flipped arrows,
obtained by applying $G_{\theta_s'}=\openone +
\theta_{s}'\sigma_x$
Two inpointing 
arrows are mapped to a singlet, and one (three)
to $\ket{\uparrow}$ ($\ket{\uparrow}^{\otimes
3}$) (Fig.~\ref{fig:rvb}c). However, the resulting ansatz does not correctly reproduce the
effect of a local field in lowest order -- breaking a
singlet into a pair of spinons -- as it 
also yields $4$-spinons-terms.  
We therefore introduce a second ansatz (``spinon pairs'') by 
replacing the singlets in
$\ket{{\bm\sigma}(D)}$ with a pair of spinons $\ket{\uparrow\uparrow}$,
i.e.\ changing each singlet to
$\tfrac{1}{\sqrt{2}}(\ket{\uparrow\downarrow}-\ket{\downarrow\uparrow})+
\theta_s^2\ket{\uparrow\uparrow}$. At the dimer point, each spinon is
tagged by a third (orthogonal) state $\ket{a_s}$ of the arrow qubit [i.e.\
$\tfrac{1}{\sqrt{2}}(\ket{\uparrow\downarrow}-\ket{\downarrow\uparrow})
\ket{a_\uparrow,a_\downarrow}+
\theta_s^2\ket{\uparrow\uparrow}\ket{a_s,a_s}$], which can be continuously
erased through a filtering $\tilde E_\lambda = 3\lambda\openone +
(1-\lambda)P$, $P_{ij}=1\,\forall\,i,j$.  Unlike the other ansatz,
 this correctly captures the expected behavior 
in leading order.  We have also found that it
performs significantly better as a variational ansatz for the Heisenberg
model with magnetic field.  We therefore focus on it,
and discuss the other ansatz in
Appendix~\ref{app:dualtension}.

\section{Results and analysis}

We will now study the response of the RVB spin liquid to different fields.
For all simulations, we have expressed the 
wavefunctions as Projected Entangled Pair States
(PEPS)~\cite{verstraete:2D-dmrg,orus:tn-review,bridgeman:interpretive-dance,schuch:rvb-kagome};
see Appendix~\ref{app:peps}.
We use standard numerical
PEPS methods (boundary
MPS~\cite{jordan:iPEPS,haegeman:medley,iqbal:z4-phasetrans})
which allow us to evaluate physical observables in
the thermodynamic limit, as well as to extract a correlation length from
the boundary MPS.  Using the techniques in
Refs.~\onlinecite{duivenvoorden:anyon-condensation,iqbal:z4-phasetrans},
we can moreover extract correlation lengths by anyon sector, which label the
decay of correlations between a pair of anyons of a certain type.  
This allows us to microscopically analyze how 
the system is driven into a trivial phase
due to doping with some anyon $a$, causing
it to condense -- in this
process, the mass gap of $a$ decreases until it eventually
vanishes and it becomes favorable to have a macroscopic number of $a$
anyons in the ground state: Anyon $a$ has become condensed,
leading to a breakdown of topological
order~\cite{bais:anyon-condensation}.  In order to probe the anyon mass
$m_a$, we can study the anyon-anyon correlation length $\xi_a\sim
1/m_a$; a diverging $\xi_a$ thus indicates condensation of anyon~$a$.

\begin{figure}[t]
\includegraphics[width=.9\columnwidth]{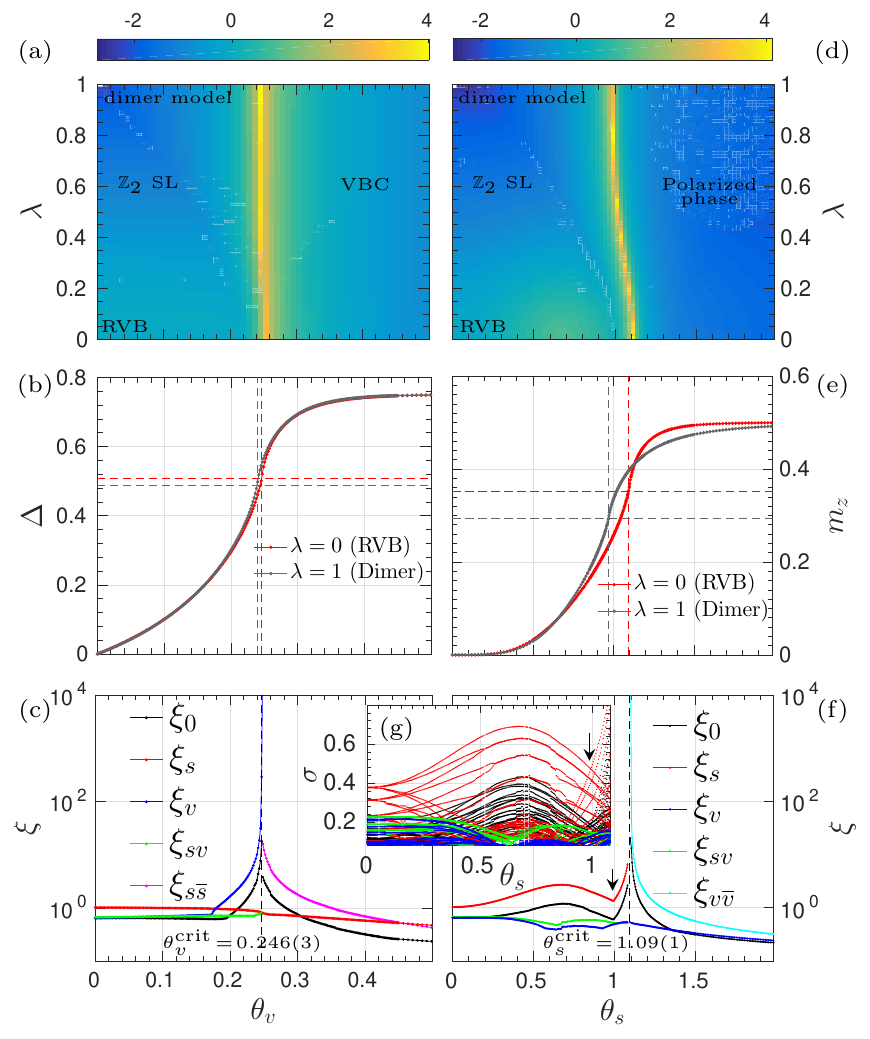}
\caption{Effect of vison doping (left) and spinon doping (right)
(joint $x$ axes).
\textbf{(a,d)} Phase diagram, driving the system into (a) a VBC phase and
(d) a spin polarized phase, respectively, where the plot shows the
logarithm of the correlation length. In both cases, the transition point
shows only weak dependence on the RVB-dimer interpolation. \textbf{(b,e)}
Response of the wavefunction to doping for RVB and dimer point, again
showing very similar behavior. \textbf{(c,f)} Anyon correlations for the
RVB as a function of doping, providing an explanation for the robustness
(see text). \textbf{(g)} Zoom into (f) including sub-leading correlations, with
$y$ axis $\sigma=e^{-1/\xi}$.}
\label{fig:allplots}
\end{figure}

We start by considering doping with visons due to lattice anisotropies which drive
the system into a vison condensed phase (i.e., a valence bond crystal
(VBC))~\cite{ralko}.
Fig.~\ref{fig:allplots}a shows the phase diagram as a function of
$\lambda$ and the anisotropy $\theta_v$.  We find that the critical point
$\theta_{v}^{\mathrm{crit}}(\lambda)$ is essentially independent of the
interpolation $\lambda$ between dimer and RVB.  Fig.~\ref{fig:allplots}b
shows the response to a doping 
$\theta_v$, defined as the difference $\Delta$ between the Heisenberg
energies on inequivalent edges, confirming that RVB and dimer model
exhibit behave essentially the same way.  We can
understand this behavior by considering the correlations (mass gaps) for
the different
anyon types in the RVB state as a function of $\theta_v$, 
Fig.~\ref{fig:allplots}c. We find that at the RVB point, the dominant
length scale is given by spinon-spinon correlations
$\xi_s$~\cite{haegeman:shadows}.  Vison
correlations $\xi_v$,  while present, are only on the order of the
topologically trivial correlations $\xi_t\approx \xi_v$, which in turn are
roughly $\xi_t\approx\xi_s/2$, and thus understood as arising from
correlations between two pairs of spinons (which are topologically trivial).
That is, the dominant length scale in the system arises from spinons,
while visons do not exhibit independent correlations on their own.
As we increase the doping $\theta_v$, genuine vison correlations start 
building up, but the overall correlation length remains dominated by
the spinon correlations, which do not respond to the vison doping; only
very close to the phase transition ($\theta_v\approx 0.18$), the
vison correlations 
 start to exceed the spinon correlations and diverge at the
phase transition.

Next, we consider the effect of magnetic fields, amounting to
doping with spinons (the ``spinon pairs'' ansatz).
Fig.~\ref{fig:allplots}d shows the phase diagram as a function of the
doping $\theta_s$ and the dimer-RVB interpolation 
$\lambda$.  Surprisingly, we find that despite the dominant spinon
correlations in the RVB state,
the phase boundary stays almost constant (the RVB is
even slightly more robust). Again, studying the response
$m_z=\langle\tfrac1N\sum \sigma_z\rangle$ vs.\
$\theta_s$, Fig.~\ref{fig:allplots}e, we find two very similar curves, 
and the RVB shows a smaller response in the relevant regime; as
expected, the phase transitions coincide with maximal
susceptibility.\footnote{Note that $m_z$ does not directly measure the
spinon density for the RVB, since non-zero contributions also arise from
pairs of spinons connected by singlets in
\protect{$\langle\psi\vert\sigma_z\vert\psi\rangle$}.}

These findings are rather counterintuitive, given the dominant spinon
correlations in the RVB.  To analyze this, we consider the correlations by
anyon type, Fig.~\ref{fig:allplots}f: We find that the spinon correlation
dominates throughout, but it decreases after an initial increase, and
exhibits a sharp kink around $\theta_s\approx0.99$ after which it diverges. To study this
further, we consider the full spectrum of correlations in
Fig.~\ref{fig:allplots}g, where we make two noteworthy observations.
First, the leading spinon correlation is 4-fold degenerate, where one
would have naively expected a spin-$\tfrac12$ doublet. Moreover, the
correlation in this quadruplet which dominates at small doping is
different from the one which finally drives the phase transition -- the two
lines exhibit a sharp crossing at $\theta_s\approx
0.99$, suggesting they are distinguished by some
symmetry.  Indeed, such a symmetry protection could explain the surprising
robustness of the RVB model, since the correlations in the sector driving the
phase transition initially \emph{decrease} under doping.

\begin{figure}[t]
\includegraphics[width=\columnwidth]{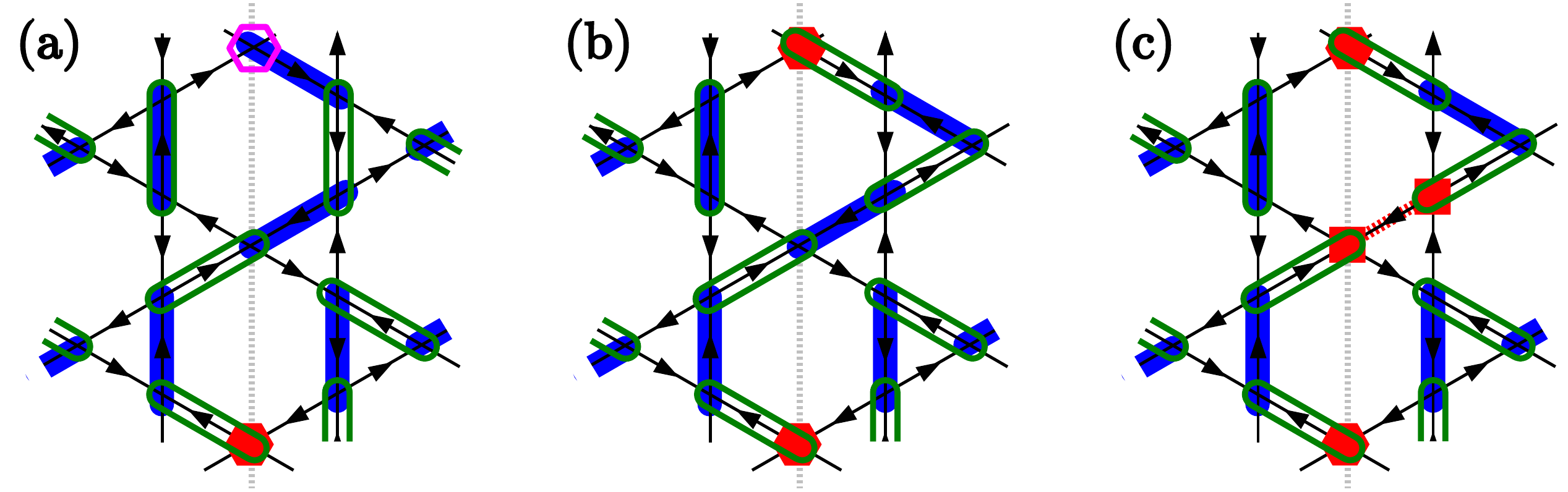}
\caption{
Spinon correlations. Figures show ket-bra overlaps (ket=full symbols,
bra=empty symbols). \textbf{(a,b)} Correlations between spinons in
(a) ket-bra and (b) ket-ket.  Each pattern has a reflection symmetric twin with
same (opposite) sign, so that only the ket-bra correlations in (a)
survive. \textbf{(c)}~Doping with pairs of spinons (red squares) inverts the sign
pattern, giving rise to ket-ket correlations.  In both cases, only
correlations of spins with same $S_z$  are possible.}
\label{fig:sp-sp-corrs}
\end{figure}

To analyze this further, we first consider the origin of spinon correlations.
Correlation functions are overlaps $\langle\psi'\vert\psi\rangle$,
where both $\ket{\psi}$ and $\bra{\psi'}$ are RVB states 
possibly doped with spinons -- that is, they are a sum of all singlet
coverings, except for one or two static locations where $\ket{\uparrow}$
spinons are placed. We will call $\ket\psi$ the ``ket layer'' and
$\bra{\psi'}$ the ``bra layer''.
Spinon-spinon correlations are obtained by summing over all overlaps of singlet patterns
with the two spinons fixed, Fig.~\ref{fig:sp-sp-corrs}, where each
pattern yields an amplitude determined by the loop lengths, and the sign
follows from the singlet orientations.  There are two types 
of such
correlations: A $\uparrow_{\mathrm{ket}}$ spinon (i.e., a single up spin
in the ket layer) can be correlated either to a $\uparrow_{\mathrm{bra}}$
spinon, Fig.~\ref{fig:sp-sp-corrs}a,
or a
$\downarrow_\mathrm{ket}$ spinon, Fig.~\ref{fig:sp-sp-corrs}b (since
$S_{z,\mathrm{total}}^\mathrm{ket}=S_{z,\mathrm{total}}^\mathrm{bra}$);
and correspondingly for a $\downarrow_{\mathrm{ket}}$ spinon. 
 We can now explain the 4-fold degeneracy: 
In the $S_z^\mathrm{total}=+\tfrac12$ sector, either a
$\uparrow_\mathrm{ket}$ spinon is correlated with
$\alpha\uparrow_\mathrm{bra}+
\beta\downarrow_\mathrm{ket}$ in a fixed superposition, or independently
$\downarrow_\mathrm{bra}$ with a superposition 
$\alpha\downarrow_\mathrm{ket}+
\beta\uparrow_\mathrm{bra}$.
This yields a 2-fold degeneracy, while another
factor of $2$ arises
from the spin-$\tfrac12$ doublet.

However, there is more structure to the spinon-spinon correlations: Every
overlap pattern in Fig.~\ref{fig:sp-sp-corrs}(a,b) has a ``twin'' under reflection
about the indicated vertical axis. For the singlet orientation chosen, all
singlets are flipped under reflection: Thus, overlaps for odd length 
paths -- which connect ket-ket and bra-bra spinons -- change 
their sign under
reflection and thus cancel:\footnote{Closed loops have even length and
thus never change their sign.} In the RVB state, only spinons
with same $S_z$ (equivalently, in opposite layers) can be correlated.
Remarkably, this property is preserved under doping
(Fig.~\ref{fig:sp-sp-corrs}b): Since spinon pairs
are symmetric under reflection and do not flip the spin, 
the paths containing an odd (even) number of spinon pairs 
which don't cancel with their reflections 
are exactly those which correlate spinons with
\emph{the same} $S_z$ 
in the same (opposite) layer.  Together with the
ket$\leftrightarrow$bra symmetry, this implies that
the spinon-spinon correlations can be labelled by sectors
$\uparrow_\pm := \uparrow_\mathrm{ket}\pm\uparrow_\mathrm{bra}$ and 
$\downarrow_\pm := \downarrow_\mathrm{ket}\pm\downarrow_\mathrm{bra}$
(that is, sectors which have overlap only with the corresponding
superposition of spinons).  At the RVB point, they all
appear with equal amplitude but opposite phase such that the equal-layer
correlations cancel. Crucially, as shown above, this symmetry label is protected by
reflection symmetry even at finite doping, preventing mixing.
Our analysis is confirmed by numerically computing the matrix elements of
$\updownarrow_\pm$ with the different correlation 
sectors.\footnote{\protect{$\updownarrow$} denotes the two cases
\protect{$\uparrow$} and
\protect{$\downarrow$} jointly, analogous to \protect{$\pm$}.}
Moreover, analysis of the data shows that the correlations which increase
initially are in the $\updownarrow_-$ sectors, while those in the
$\updownarrow_+$ sectors decrease, whereas the phase
transition is driven by a diverging correlation length in the $\uparrow_+$
sector.  However, coupling between the two sectors is prohibited
by reflection symmetry. This explains why the system responds to spinon
doping with an increased correlation length, and yet, this does not come
with a decrease in robustness of the topological phase: the two phenomena
take place in different symmetry sectors.
A qualitatively similar behavior is observed for the ``dual tension'' doping:
The spinon correlation spectrum again splits in the 
$\updownarrow_\pm$ basis, and 
we find only a very weak effect on the robustness.

While our analysis is based on a specific doping model, this is in
fact a universal behavior, since any perturbation which results in a
breaking of singlets into pairs of spinons will have the same effect to
leading order.  This implies that the initial splitting will again be in
the $\updownarrow_\pm$ basis, with the $\updownarrow_-$ correlations being
dominant for small doping, while the $\uparrow_+$ correlations drive the
phase transition.  As long as the perturbation respects the lattice
symmetry, these correlation sectors cannot mix, and we therefore expect a
qualitatively similar behavior for a general perturbation which induces
doping with spinons.

\section{Conclusions}

We have studied the robustness of the RVB and the dimer model on the
kagome lattice to perturbations using PEPS. We have found that despite the
non-orthogonality of different singlet configurations, the RVB spin liquid
exhibits the same robustness to perturbations as the dimer model. For
lattice anisotropies (doping with visons), we traced this back to the fact
that the length scale induced by the non-orthogonality of singlets gives
rise to spinon correlations but does not directly affect the physics of
visons. For magnetic fields (doping with spinons), we showed that the
robustness arises from a protection of the RVB state due to the reflection
symmetry of the lattice, which separates the initially dominating spinon
branch from the branch which ultimately drives the phase transition. Our
results reveal a surprising universal robustness of the RVB spin liquid
against perturbations, highlighting its role as a candidate for the
realization of a stable gapped spin liquid.

\acknowledgements
This work has received support by the European Research Council (ERC)
under the European Union's Horizon 2020 Research and Innovation Programme
through the ERC Starting Grant WASCOSYS (No.~636201), and by the Deutsche
Forschungsgemeinschaft (DFG) under Germany's Excellence Strategy (EXC-2111
-- 390814868). HC acknowledges support by the International Internship
Program of Princeton University.

\onecolumngrid

\begin{center}
\vspace*{0.3cm}
\rule{12cm}{.5pt}
\vspace*{0.3cm}
\end{center}

\twocolumngrid

\newcommand{\cbox}[2]{\vcenter{\hbox{\includegraphics[width=#1em]{#2}}}}
\newcommand{\dbox}[3]{\vcenter{\hbox{\includegraphics[width=#1em,height=#2em]{#3}}}}

\appendix

\section{Tensor network implementation of doped RVB\label{app:peps}}

Here, we describe how the different doping mechanisms introduced in the
paper can be described as tensor network states, also termed Projected
Entangled Pair States (PEPS).

\subsection{String tension and dual tension}

Let us first describe the PEPS for the RVB with vison and spinon dopings
constructed from doping of the underlying loop (or arrow) model, i.e.,
string tension or dual string tension.
 This construction will consist of three 
layers, stacked on top of each other.  The lower layer is a PEPS for the
loop model with the corresponding (dual) tension.  On top of that layer,
we apply a PEPO (Projected Entangled Pair Operator) which transforms this
loop model into the corresponding dimer model. Finally, in a last step we
apply local filtering operations (as introduced in the main text) to
the arrow degrees of freedom in the dimer model, which allows to
interpolate to the corresponding RVB state.

\begin{figure}[t]
	\centering
	\includegraphics[width=20em]{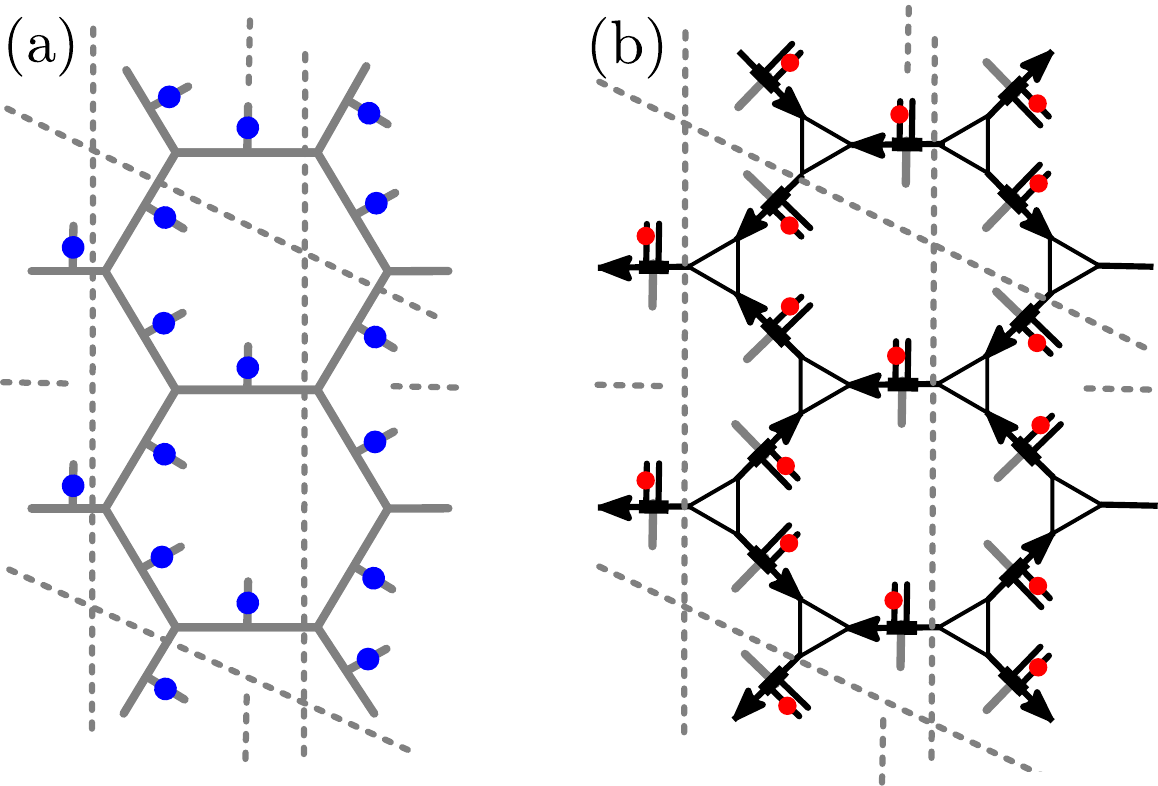}
	\caption{\textbf{(a)} A tensor network of the loop model on the honeycomb
lattice with filtering (blue bubble) on each site. \textbf{(b)}~PEPO for mapping
the loop model to a dimer or RVB model.  The on-site tensors are oriented
as prescribed by the reference arrow pattern. The incoming (gray) indices
are contracted with physical indices in panel (a). } \label{fig:tn}
\end{figure}

The PEPS for the first layer -- the loop model with tension -- consists of
two types of tensors: One vertex tensor (without physical legs) and an
on-site tensor (which carries the physical index). The vertex tensor has
three legs, each of dimension two. We use a computational basis to express
the presence or absence of a loop string on the link, and due to the
$\mathbb{Z}_2$ constraint, the vertex tensor is restricted to four
non-zero entries.  
\begin{equation}
\begin{split}
\cbox{2.5}{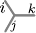} =\ & \delta_{i0}\delta_{jk1} + \delta_{j0}\delta_{ik1} +  \delta_{k0}\delta_{ij1} +  \delta_{ijk0}.
\end{split}
\end{equation}
We use $\delta_{i_1i_2..i_n}$ to denote an entry of the $\delta$ tensor with $n$ indices, and the entry is one iff all the indices are equal. The on-site tensor on every link syncs up indices of adjoining vertex tensors and the physical index of the loop model.
\begin{equation}\label{eq:tc_P}
\cbox{2.7}{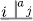} = \delta_{ija}
\end{equation}
After blocking the vertex and the on-site tensors \eqref{eq:tc_P}, we get
the tensor network of the loop model on an infinite lattice
(Fig.~\ref{fig:tn}a). We can filter different loop configurations in the
diagonal and dual basis by applying a deformation of the form
\begin{equation}\label{eq:str_ten}
\cbox{2.0}{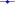} = \begin{pmatrix} 
1 & \theta_s' \\
\theta_s' & 1-\theta_v
\end{pmatrix}
\end{equation}
on physical legs. (In the main text, we restrict to the cases where one of the
$\theta_\bullet\equiv0$).

The second layer is a PEPO which maps loop configurations 
on the honeycomb lattice (including broken loops in the case of dual tension)
to dimer configurations on the kagome lattice (including monomers in
the case of broken loops). It again consists of two types of tensors.  The
first is a triangular tensor without physical index, which has 
three indices of dimension three each:
\begin{equation}
\label{eq:A4}
\begin{split}
\cbox{2.4}{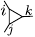} =\ & \sigma_{ij} \delta_{k2} + \sigma_{jk} \delta_{i2} + \sigma_{ki} \delta_{j2} +  \delta_{ijk2}+ \\ &
\delta_{i0}\delta_{jk2} + \delta_{j0}\delta_{ik2} +
\delta_{k0}\delta_{ij2} + \delta_{ijk0}\ ,
\end{split}
\end{equation}
where the tensor $\sigma = \text{diag} ( \frac{1}{\sqrt{2}} 
\begin{psmallmatrix*}[r]0 & -1\\ 1 & 0\end{psmallmatrix*}, 0)$ 
models an oriented singlet and the last four terms correspond to different
spinon configurations. The second tensor acts on each site: It takes the
loop configuration as an input in index $a$, and outputs a physical spin
$p$ and an arrow index $d$, and is built such as to pick the physical qubit
from either of the two adjacent virtual indices (and thus triangular
tensors \eqref{eq:A4}), as prescribed by the reference configuration:
\begin{equation}\label{eq:P}
\begin{split}
\cbox{3.0}{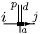} =\ \delta_{ad}(& \delta_{d0}\delta_{j2}(\delta_{i0}\delta_{p0} + \delta_{i1}\delta_{p1} ) +
\\ & \delta_{d1}\delta_{i2}(\delta_{j0}\delta_{p0} + \delta_{j1}\delta_{p1} ))
\ .
\end{split}
\end{equation}
By design, this tensor is not symmetric in the virtual indices $i$ and
$j$, and we use an arrow pointing to the index $j$ to label its
orientation. The PEPO (Fig.~\ref{fig:tn}b) is now obtained by assembling
\eqref{eq:A4} and \eqref{eq:P} in a hexagonal structure (yielding spins on
a kagome lattice), where the arrows need to be oriented such that setting
$a\equiv 0$ yields the reference configuration.  

The tensor network for the doped dimer model is now obtained by stacking
the two tensor networks Fig.~\ref{fig:tn}a (for the honeycomb loop model)
and Fig.~\ref{fig:tn}b (for replacing loops with dimers), where the
gray indices (labelled $a$) are contracted.
The resulting tensor network allows to tune the doping with
spinons and visons by changing the parameters 
$\theta_s'$ and $\theta_v$ in the 
deformation tensor \eqref{eq:str_ten}.

Finally, by applying a filtering 
\begin{equation}
\cbox{2.0}{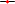} \equiv E_\lambda =  \begin{pmatrix} 1+\lambda & 1-\lambda \\ 1-\lambda
& 1+\lambda \end{pmatrix} \end{equation}
on the arrow qubits $d$, we can continuously interpolate between the dimer
and RVB models also with doping.

\subsection{Doping with spinon pairs}
The tensor network to implement the ``spinon pair'' doping 
is obtained by modifying the second layer of the preceding construction.
There is no longer a need for the first layer (the loop model), since the
loop constraint is already contained in the dimer model (see also the
original construction for the RVB and dimer PEPS~\cite{schuch:rvb-kagome}).
First, the tensor \eqref{eq:A4} is modified such that each
index has dimension four:
\begin{equation}
\cbox{2.4}{eps} = \sigma_{ij} \delta_{k2} + \sigma_{jk} \delta_{i2} + \sigma_{ki} \delta_{j2} +  \delta_{ijk2}.
\end{equation}
Here, the tensor $\sigma = \text{diag} ( \frac{1}{\sqrt{2}}
\begin{psmallmatrix*}[r]0 & -1\\ 1 & 0\end{psmallmatrix*}, 0,1)$ encodes
either a singlet (in the first two basis states) or the presence of 
a spinon pair (in the new fourth degree of freedom).
Correspondingly, the on-site tensor is also changed to project to the spinon
degree of freedom with a tunable weight of $\theta_s$, accompanied by a
third state $d=2$ of the arrow qubit (the basis state $\ket{a_s}$):
\begin{equation}
\begin{split}
\cbox{2.4}{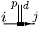} =\ & 
    \delta_{d0}\delta_{j2}( \delta_{i0}\delta_{p0} + \delta_{i1}\delta_{p1}) + \\
  & \delta_{d1}\delta_{i2}( \delta_{j0}\delta_{p0} + \delta_{j1}\delta_{p1}) + \\
  & \theta_s \delta_{d2}\delta_{p0}(\delta_{i3}\delta_{j2}+\delta_{i2}\delta_{j3})\ ,
\end{split}
\end{equation}
where the arrows are oriented as before.
In order to interpolate to the RVB state, we can erase the information on
the dimer indices by applying a deformation 
\begin{equation}
\cbox{2.0}{dimer_ten} \equiv
\tilde E_\lambda = 
\begin{psmallmatrix*}[r]
1+2\lambda & 1-\lambda &1-\lambda\\
1-\lambda & 1+2\lambda &1-\lambda\\
1-\lambda & 1-\lambda &1+2\lambda
\end{psmallmatrix*}\ .
\end{equation}
The final tensor network is identical to Fig.~\ref{fig:tn}b, but without
the gray indices.

\section{spinon doping with ``dual tension''\label{app:dualtension}}

In this appendix, we report the results for the model with spinon doping
constructed through dual string tension.

\begin{figure*}[t]
\parbox{.98\columnwidth}{
\includegraphics[width=.98\columnwidth]{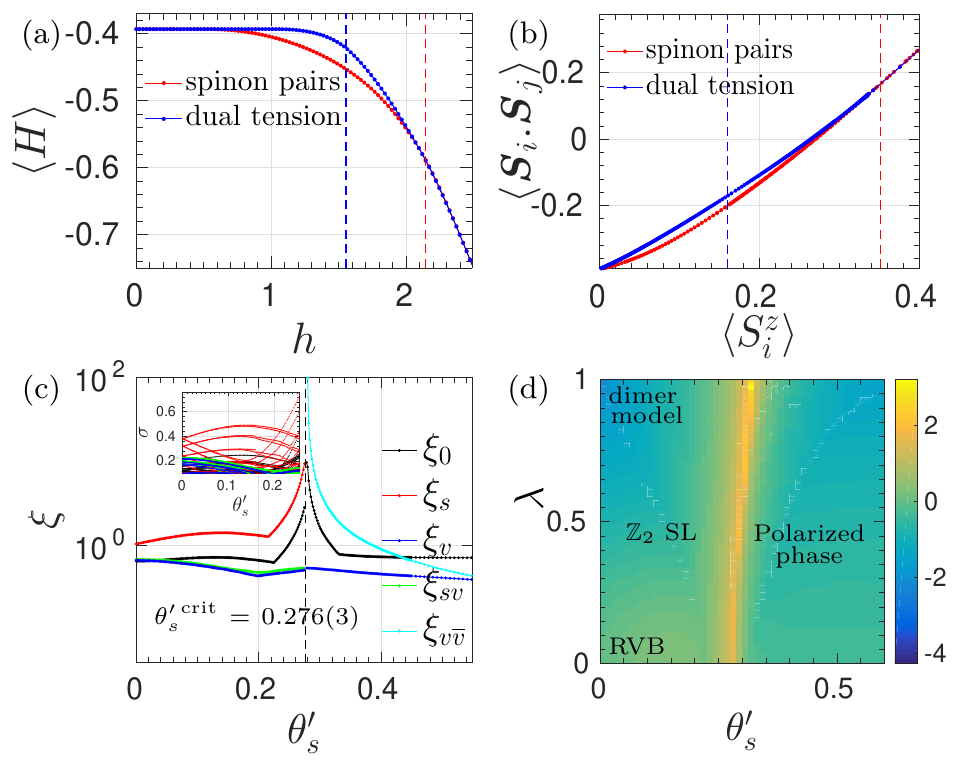}
}
\hspace*{\columnsep}
\parbox{.98\columnwidth}{
\caption{Spinon doping: ``dual tension'' model. \textbf{(a)} Comparison of
variational energies for ``spinon pairs'' and ``dual tension'' ansatz as a
variational wavefunction for the Heisenberg model with field,
Eq.~\eqref{eq:app:heis-field}; the ``spinon pairs'' ansatz performs clearly better
for most of the parameter regime. 
Here and in (b), the dashed lines indicate the respective phase
transitions.
\textbf{(b)} Comparison of Heisenberg energy
vs.\ magnetization for the two models. This illustrates that the two
models already differ in the perturbative regime (close to the origin);
only the ``spinon pairs'' ansatz correctly captures the physics of breaking a
singlet into a nearest-neighbor pair of spinons in leading order.
\textbf{(c)} Correlation functions by spinon sector for the ``dual
tension'' spinon doping; we observe the same characteristic features as
in Fig.~\ref{fig:allplots}f, which are protected by the same symmetries
(in particular lattice reflection) as discussed in the main text.
\textbf{(d)} The phase diagram of the dimer-RVB interpolation $\theta_s'$
for the ``dual tension'' doping $\lambda$ again exhibits only a weak dependence of
the phase transition on changing orthogonal dimers to non-orthogonal
singlets.
}
\label{fig:app-dual}
}
\end{figure*}

Fig.~\ref{fig:app-dual}a provides a comparison of the variational energy
for the ``dual tension'' and the ``spinon pairs'' ansatz for the
Heisenberg Hamiltonian with a transverse field,
\begin{equation}
\label{eq:app:heis-field}
H=\sum_{\langle ij\rangle} \bm{S}_i\cdot\bm{S}_j - h\sum_i S_i^z
\end{equation}
(with eigenvalues $S_i^z=\pm \tfrac12$). 
We find that the energy for the ``spinon pairs'' ansatz is significantly
lower, providing a first reason why we chose to consider it as our primary
ansatz for spinon doping. Fig.~\ref{fig:app-dual}b provides further
insight into this.  It shows the relation between Heisenberg energy 
$\langle\bm{S}_i\cdot\bm{S}_j\rangle$ and magnetization $\langle
S_i^z\rangle$: We see that for the same magnetization, the ``dual
tension'' ansatz has a significantly higher Heisenberg energy, which 
qualitatively means that it requires to break up a correspondingly larger number
of singlets to achieve the same magnetization. This effect can be clearly
observed in the perturbative regime, i.e., small magnetizations, where the
``dual tension'' ansatz has a significantly higher slope.  We can
understand this effect qualitatively in a semi-classical picture: Breaking
up a singlet into a pair of spinons leads to a change $\Delta E=1$ in
Heisenberg energy, since one singlet is replaced by a
$\left\vert\uparrow\uparrow\right\rangle$ state. On the other hand, the scenario
in the ``dual tension'' construction
where flipping an arrow yields $4$ spinons ($3$ on a
triangle, and one adjacent vertex) gives rise to a total of four
Heisenberg terms having an energy $+\frac 14$, while before, half of them
had energy $0$ and half $-\tfrac34$, implying $\Delta E=2.5$ (or $1.25$
per pair of spinons).

Despite these differences, the study of correlations by anyon sectors,
Fig.~\ref{fig:app-dual}c, yields a qualitatively very similar behavior to
the case of ``spinon pair'' doping.  In particular, we again observe an
additional two-fold degeneracy in the spinon sector (on top of the
spin-$\tfrac12$ multiplet) which is protected by the lattice symmetry, and
which separates the correlations responsible for the phase transitions
from those initially responding to the doping; this highlights the fact
that this, as we have shown, is a universal effect.  Yet again, this
symmetry protection is reflected in a rather weak dependence of the phase
transition on interpolating between the doped orthogonal dimer model and the
doped RVB state, Fig.~\ref{fig:app-dual}d.
\\ \quad


\begin{thebibliography}{10}

\bibitem{balents:spin-liquid-review-2010}
L.~Balents,
\newblock {\em Spin liquids in frustrated magnets},
\newblock Nature {\bf 464}, 199 (2010).

\bibitem{savary:spin-liquids}
L.~Savary and L.~Balents,
\newblock {\em Quantum Spin Liquids},
\newblock Rep. Prog. Phys. {\bf 80}, 016502 (2017), arXiv:1601.03742.

\bibitem{knolle:spin-liquids}
J.~Knolle and R.~Moessner,
\newblock {\em A Field Guide to Spin Liquids},
\newblock Annu. Rev. Condens. Matter Phys. {\bf 10}, 451 (2019),
  arXiv:1804.02037.

\bibitem{anderson:rvb}
P.~W. Anderson,
\newblock {\em Resonating valence bonds: A new kind of insulator?},
\newblock Mater. Res. Bull. {\bf 8}, 153 (1973).

\bibitem{misguich:AFM-review}
G.~Misguich and C.~Lhuillier,
\newblock {\em Two-dimensional quantum antiferromagnets},
\newblock in {\em Frustrated spin systems}, edited by H.~T. Diep,
  World-Scientific, 2003, cond-mat/0310405.

\bibitem{schuch:rvb-kagome}
N.~Schuch, D.~Poilblanc, J.~I. Cirac, and D.~P{\'e}rez-Garc{\'\i}a,
\newblock {\em Resonating valence bond states in the PEPS formalism},
\newblock Phys. Rev. B {\bf 86}, 115108 (2012), arXiv:1203.4816.

\bibitem{zhou:rvb-parent-onestar}
Z.~Zhou, J.~Wildeboer, and A.~Seidel,
\newblock {\em Ground state uniqueness of the twelve site RVB spin-liquid
  parent Hamiltonian on the kagome lattice},
\newblock Phys. Rev. B {\bf 89}, 035123 (2014), arXiv:1310.8000.

\bibitem{rokhsar:dimer-models}
D.~S. Rokhsar and S.~A. Kivelson,
\newblock {\em Superconductivity and the Quantum Hard-Core Dimer Gas},
\newblock Phys. Rev. Lett. {\bf 61}, 2376 (1988).

\bibitem{misguich:dimer-kagome}
G.~Misguich, D.~Serban, and V.~Pasquier,
\newblock {\em Quantum Dimer Model on the Kagome Lattice: Solvable Dimer-Liquid
  and Ising Gauge Theory},
\newblock Phys. Rev. Lett. {\bf 89}, 137202 (2002), cond-mat/0204428.

\bibitem{elser:rvb-arrow-representation}
V.~Elser and C.~Zeng,
\newblock {\em kagome spin-1/2 antiferromagnets in the hyperbolic plane},
\newblock Phys. Rev. B {\bf 48}, 13647 (1993).

\bibitem{kitaev:toriccode}
A.~Kitaev,
\newblock {\em Fault-tolerant quantum computation by anyons},
\newblock Ann. Phys. {\bf 303}, 2 (2003), quant-ph/9707021.

\bibitem{verstraete:2D-dmrg}
F.~Verstraete and J.~I. Cirac,
\newblock {\em Renormalization algorithms for Quantum-Many Body Systems in two
  and higher dimensions},
\newblock (2004), cond-mat/0407066.

\bibitem{orus:tn-review}
R.~Orus,
\newblock {\em A Practical Introduction to Tensor Networks: Matrix Product
  States and Projected Entangled Pair States},
\newblock Ann. Phys. {\bf 349}, 117 (2014), arXiv:1306.2164.

\bibitem{bridgeman:interpretive-dance}
J.~C. Bridgeman and C.~T. Chubb,
\newblock {\em Hand-waving and Interpretive Dance: An Introductory Course on
  Tensor Networks},
\newblock J. Phys. A: Math. Theor. {\bf 50}, 223001 (2017), arXiv:1603.03039.

\bibitem{jordan:iPEPS}
J.~Jordan, R.~Orus, G.~Vidal, F.~Verstraete, and J.~I. Cirac,
\newblock {\em Classical simulation of infinite-size quantum lattice systems in
  two spatial dimensions},
\newblock Phys.\ Rev.\ Lett. {\bf 101}, 250602 (2008), cond-mat/0703788.

\bibitem{haegeman:medley}
J.~{Haegeman} and F.~{Verstraete},
\newblock {\em {Diagonalizing Transfer Matrices and Matrix Product Operators: A
  Medley of Exact and Computational Methods}},
\newblock Annual Review of Condensed Matter Physics {\bf 8}, 355 (2017),
  arXiv:1611.08519.

\bibitem{iqbal:z4-phasetrans}
M.~Iqbal, K.~Duivenvoorden, and N.~Schuch,
\newblock {\em Study of anyon condensation and topological phase transitions
  from a Z4 topological phase using Projected Entangled Pair States},
\newblock Phys. Rev. B {\bf 97}, 195124 (2018), arXiv:1712.04021.

\bibitem{duivenvoorden:anyon-condensation}
K.~{Duivenvoorden}, M.~{Iqbal}, J.~{Haegeman}, F.~{Verstraete}, and
  N.~{Schuch},
\newblock {\em {Entanglement phases as holographic duals of anyon
  condensates}},
\newblock Phys. Rev. B {\bf 95}, 235119 (2017), 1702.08469.

\bibitem{bais:anyon-condensation}
F.~Bais and J.~Slingerland,
\newblock {\em Condensate induced transitions between topologically ordered
  phases},
\newblock Phys. Rev. B {\bf 79}, 045316 (2009), arXiv:0808.0627.

\bibitem{haegeman:shadows}
J.~Haegeman, V.~Zauner, N.~Schuch, and F.~Verstraete,
\newblock {\em Shadows of anyons and the entanglement structure of topological
  phases},
\newblock Nature Comm. {\bf 6}, 8284 (2015), arXiv:1410.5443.

%
%
%
\bibitem{ralko}
A.~Ralko, M.~Ferrero, F.~Becca, D.~Ivanov, and F.~Mila,
\newblock{\em{Crystallization of the resonating valence bond liquid as
vortex condensation,}}
Phys. Rev. B \textbf{76}, 140404(R).
\end{thebibliography}
\end{document}